\documentstyle[12pt]{article}
\begin{document}

\newcommand{\newc}{\newcommand}
\newc{\beq}    {\begin{equation}}
\newc{\eeq}    {\end{equation}}
\newc{\beqa}    {\begin{eqnarray}}
\newc{\eeqa}    {\end{eqnarray}}
\newc{\ba}    {\begin{array}}
\newc{\ea}    {\end{array}}
\newc{\st}    {\stackrel}
\newc{\f}    {\frac}

\def\NCA{\em Nuovo Cimento}
\def\NPB{{\em Nucl. Phys.} {\bf B}}
\def\PLB{{\em Phys. Lett.}  {\bf B}}
\def\PRL{{\em Phys. Rev. Lett. }}
\def\PRD{{\em Phys. Rev.} {\bf D} }
\def\PRP{{\it  Phys. Rep.} }
\def\ZPC{{\em Z. Phys.} {\bf C}}

\title{RUNNING INFLATION }
\author{ Jae-weon Lee,\\
      \it Department of Physics,             
      \\ \it          Hanyang University,
                Seoul, 133-791, Korea
        \\          
 and\\
 In-gyu Koh \\ \it Department of Physics,
 \\ \it Korea Advanced Institute of Science and Technology,
\\   \it 373-1, Kusung-dong, Yusung-ku, Taejeon, Korea  \\
\\
}

\date{}
\maketitle

\begin{abstract}

 A first order inflation model where  a gauge
coupling constant runs
 as the universe inflates is investigated.
 This model can solve the graceful-exit problem
 within Einstein gravity by varying
the bubble formation rate.
The  sufficient expansion condition
gives group theoretical constraints on the inflaton field,
while the appropriate density perturbation requires an additional
scalar field or cosmic strings.
 
 \end{abstract}
\maketitle


Keywords: inflation, coupling constant, bubble,
   renormalization group, density perturbation, cosmic string
\newpage

  It is well known that inflation\cite{old} could solve many problems
of the standard big-bang cosmology such
 as the flatness and the horizon problems.
However, from the very beginning, it have been
known that the original inflation model(old inflation)
 has its own problem, so called `graceful-exit' problem. 

For a sufficient expansion to solve the problems of the big-bang cosmology
 a small bubble
formation rate is required, while to complete the
phase transition a large rate is required.
It was shown that these two condition can  not be satisfied, 
simultaneously\cite{guth}.

  To avoid the  troublesome bubble formations,
another model, so-called the `new' inflation model was 
introduced\cite{slowroll}.
However, this model has fine-tuning problem to explain the small
density perturbation and a sufficient expansion.
It is also difficult to establish an initial thermal equilibrium 
required for the inflation.
On the other hand, chaotic inflation model\cite{chaotic} uses  
this non-equilibrium state  to give the initial conditions
for the inflaton fields.

A few years ago,
 the first-order inflation model with  Brans-Dicke gravity
(extended inflation model) 
was suggested to solve the graceful-exit problem
by varying the Hubble parameter $H$\cite{extended},
while it also has another problem, `big bubble' problem.
Even though the extended gravity sectors are favorable
in many particle physics theories, it may be still meaningful to
study the inflation models within Einstein gravity which is
empirically well supported. 

 In this paper, to overcome the graceful-exit problem with
Einstein gravity, we study the first-order inflation model,
called "running inflation" by the authors,
where the gauge coupling constant $g$  varies
 as the the universe expands exponentially.

During the inflation  
the universe should expand by the enormous factor of $z > e^{60}$,
and the typical energy scale $Q$ of the particles in the
universe should be reduced by the
factor.(After some expansion the universe enters
into non-equilibrium states. Hence, it is
more adequate to use $Q$ rather than the temperature
which could not be defined in a non-equilibrium state.)

 Hence, it is very natural to consider the running of the couplings. 
In fact, even around  the birth of the old inflation model
there are several works considering this running coupling effect
\cite{rginflation,rgimproved}.   
There, it had been shown that the running of $g$
enhances the bubble formation by the thermal fluctuation, 
and makes the phase transition duration short.
The thermal transition rate is proportional to
$exp(-S_3(T)/T)$, where $S_3(T)$ is the Euclidean
action for the $O(3)$ symmetric bubble solution
at the temperature $T$.
 
In this paper, we will consider the bubble formation 
by the quantum fluctuation rather than the thermal fluctuation. 

This is justified by the fact that in thin-wall limit as $T$ decreases,
$S_3(T)/T$ rapidly reaches its minimum and then increases 
again, therefore, the thermal transition rate becomes negligible
below the sufficiently small temperature \cite{thermal}. 

Let us first  review the graceful-exit problem. 
The crucial quantity for the bubble formation rate is
\cite{size}
\beq
\epsilon\equiv\frac{\Gamma}{H^4},
\eeq
which is the bubble nucleation rate within a Hubble volume($H^{-3}$) in 
a Hubble time ($H^{-1}$).
For a sufficient expansion it is required that  
 $\epsilon \st{<}{\sim} 4 \times 10^{-3}$, while for the
phase transition completion $\epsilon \geq \epsilon_{cr}$,
 where $ 10^{-6} \st{<}{\sim} \epsilon_{cr} \st{<}{\sim} 0.24 $
\cite{guth}.
To overcome this problem we study the change of  $\Gamma$ due to the running of the couplings.
\bigskip

 Consider the inflaton field $\phi$ which has a gauge interaction
with coupling $g$ and  the zero temperature one-loop effective 
potential \cite{potential} 
\beq
V(\phi)= (2A-B)\sigma^2 \phi^2-A\phi^4+B\phi^4 ln(\phi^2/\sigma^2)+V_0,
\label{V}
\eeq

where $V_0$ is added to set the cosmological constants to zero. 
The field $\phi$ may be a GUT (Grand unified theory) Higgs scalar.
$A$ is a free parameter
and $B= g^4 N_{deg}/64\pi^2$, 
where $N_{deg}$ is the degree of  the freedom of the particles
 interacting with $\phi$.
Then, the quantum tunneling rate of bubble  is given by
 $\Gamma\simeq T_{inf}^4 exp(-S_{4})$, where $S_{4}$ is the Euclidean 
action for the $O(4)$ symmetric bubble solution\cite{coleman}. 
Here, $T_{inf}\equiv (V_0)^{1/4}$ is chosen for the energy scale of
the inflation. 
The exponential part is sensitive to $Q$
and more important than
the prefactor for our study.

During the inflation, $\epsilon$ is related to $S_{4}$ by
\beq
\epsilon(Q) \simeq \frac{T_{inf}^4 e^{-S_{4}(Q)}}
{(\frac{T_{inf}^4}{3M^2})^2}\simeq (\frac{M}{T_{inf}})^4
e^{-S_{4}(Q)},
\label{epsilon}
\eeq
where $M=M_P/\sqrt{8\pi}$ is the reduced Planck mass.

Therefore, to solve the graceful-exit problem,
it is  required that  $S_{4}(Q)$ should satisfy the following
two conditions.

First,$\epsilon(Q_{end})\ge \epsilon_{cr}$, or,
\beq
S_4(Q_{end})\st{<}{\sim} 4ln( M/T_{inf}) -ln\epsilon_{cr},
\label{condition1}
\eeq
where $Q_{end}$ is the value of $Q$ at the end of the inflation.

Second
\beq
\frac{\epsilon(Q_{end})}{\epsilon(T_{inf})}= 
exp[S_{4}(T_{inf})-S_{4}(Q_{end})]
>\frac{\epsilon_{cr}}{4\times 10^{-3}} \simeq 60,
\label{condition2}
\eeq
 where $\epsilon_{cr}$ is set to 0.24 at the last
equality.

 The full numerical solution of $S_{4}$ for $V(\phi)$ are known
\cite{se4}. 
In thin-wall  limit 
\beq
 S_{4}=\frac{27\pi^2 I^4 B^2}{8 (A+B)^3},
\eeq
 where the number $I\simeq 0.42$.

Since $A\simeq B $ in this limit,  one can find  that 
$S_{4}\sim B^{-1} \sim g^{-4}$. 

Hence, if $g$ grows during the inflation,
$B$ increases,  $S_{4}$ decreases, and it makes $\epsilon$ large
enough to solve the graceful-exit problem.

Let us find the group theoretical constraints on $\phi$.
To investigate the energy scale dependence of the coupling 
constant $\alpha\equiv g^2/4\pi$,
we must know the symmetry group $G$ of $\phi$.
The well known solution for the one-loop 
renormalization group(RG) equation for $\alpha$ is written as\cite{RG}
\beq
\alpha(Q)=\frac{\alpha(T_{inf})}{1- \alpha(T_{inf}) b(G) t(Q)}
=\frac{1}{b(G)ln\frac{Q}{\Lambda}},
\label{alpha}
\eeq
where $t(Q)\equiv ln (T_{inf}/Q)$,
and $\Lambda\equiv T_{inf} exp(-1/\alpha(T_{inf}) b(G))$.

When $Q$ approaches to  the energy scale $\Lambda$,
$\alpha$ diverges and the perturbation theory fails. 

Here  the group $(G)$ and the representation ($R_f,R_s$) dependent factor 
\beq
b(G)=\frac{1}{6\pi}[11 C_2(G) - 2N_fT(R_f)
-N_sT(R_s)],
\label{b}
\eeq
where $N_f$ and $N_s$ are  the number of the fermions and
the  scalars interacting with $\phi$, respectively.
For a typical model with $G=SU(N)$, $ C_2(G)=N$ and $T(R_f)=1/2$.

For a supersymmetric theory
\beq
b(G)=\frac{1}{2\pi}[3 C_2(G) - \displaystyle\sum_{R} T(R)],
\label{bSUSY}
\eeq
where the sum is over fermion representations $R$.

It is demanded  that the inflation ends before the perturbation fails,
because as $Q$ approaches to $\Lambda$,
 $\alpha$ becomes large enough to make  
$\epsilon$ bigger than $\epsilon_{cr}$.
In other words, if $g$ approaches to 1, then  $S_{4}\simeq 1$,
and the first condition (eq.(\ref{condition1}))
can be easily satisfied.
(However, this is not a leading order treatment.
We will discuss this point later.)

Since $ Q $ is inversely proportional to the scale factor
 $R$, the above mentioned condition, $\Lambda<Q_{end}$,
 corresponds to 
\beq
\frac{T_{inf}}{\Lambda}>\frac{R(Q_{end})}{R(T_{inf})}\equiv z\equiv e^N.
\label{z}
\eeq
Since $T_{inf}\st{<}{\sim} 10^{16}~GeV$ by the 
density perturbation limit\cite{energy}, 
 one can find $ \Lambda \st{<}{\sim}
 10^{-10}~GeV$.

>From eq.(\ref{alpha}) and eq.(\ref{z}) it follows that 
\beq
b(G)<[\alpha(T_{inf}) ln \frac{T_{inf}}{\Lambda}]^{-1}
<\frac{1}{\alpha(T_{inf})N}.
\label{constraint}
\eeq

Since the generic GUTs favor
 $\alpha\simeq 10^{-2}$,
 and  a sufficient inflation needs 
$N\st{>}{\sim}60$, it is required that 
\beq
0<b(G)\st{<}{\sim} 1.
\label{result}
\eeq
Here, the condition $b(G)>0$ is added to
ensure that our theory should be asymptotically free to make
$\alpha(Q)$ be a decreasing function of $Q$. 

The condition for $b(G)$ (eq.(\ref{result})) seems to be
satisfied only when the gauge and the matter parts almost cancel
each other. In this respect the supersymmetric model (eq.(\ref{bSUSY}))
 would be preferred to the non-supersymmetric one (eq.(\ref{b})). 

However,  the direct inclusion of 
running coupling effect( eq.(\ref{alpha}))  into eq.(\ref{V})
as above is not leading (one-loop) order treatments\cite{rgimproved}.
The more appropriate treatment needs 
RG improved 
action for the tunneling,  
which is given by\cite{kuwari}
\beq
S_4(g(t))=e^{-N_\gamma (t)}S_4(g_0),
\label{s4improve}
\eeq
where 
$N_\gamma (t)=4\int^{t}_{0} \gamma(g(t'))dt'$,
$t(Q)\equiv ln (T_{inf}/Q)$, $g_0\equiv g(T_{inf})$,
 and $\gamma(t)$ is an anomalous dimension.

Then,
\beq
\frac{dlnS_4}{dt}=4\gamma(t).
\eeq
So if $\gamma(t)>0$, $S_4$ decreases as $Q$ decreases.     
In this case the conditions to solve the grace-exit problem (eq.\ref{condition2}) 
reduces to
\beq
exp(S_4(g_0)[1-e^{-N_\gamma(N)}])\st{>}{\sim} 60,
\eeq
where we have used the fact that $t(Q_{end})=N$. 

Generally $\gamma(t)\simeq A  g^2(t) $,
 where $A\sim 10^{-2}$ is a coefficient depending on the group $G$.  
Then, with eq.(\ref{alpha}) the result of integration is 

\beq
N_\gamma(N)=-\frac{16\pi A}{b(G)} ln(1-\alpha(T_{inf})b(G) N).
\eeq
 One can easily find that
the condition $ 0< 1-\alpha(T_{inf}) b(G) N \le 1$ , which can be
reduced to eq.(\ref{constraint}), is derived.

However, another condition $b(G)\ll 1$ is
required to make $N_\gamma(N)$ large enough to complete the phase transition,
 because the log term contributes at most $-O(1)$ for perturbative region.
Furthermore, the condition $b(g)\ll 1$ is just a sufficient condition
for a successful inflation, contrary to the naive approach.
 In other words, there is a possibility that non-perturbative effects
play  roles.

It is easy to avoid big-bubble problem for models
with varying $\Gamma$.
The following condition is enough.
\beq
S_{55}\st{>}{\sim} 4ln\frac{M}{T_{inf}}+11.5,
\eeq
where $S_{55}$ is the value of $S_4$ at $55$
e-foldings before the end of inflation\cite{big}.

Our inflation model is natural in three aspects.
First, it adopts running coupling effect
which seems to be unavoidable, if we  consider
the enormous expansion of the universe during inflation. 
Second, empirically 
well supported Einstein gravity is used.
Finally, the inflaton potential
does not need fine-tuning required for
the flat potential for new-inflation models.
 
Although our model is simple and natural,
it has a problem.
The density perturbation is the most
 stringent constraint on the inflation models.
COBE observation\cite{smoot} requires 
 the density perturbation $\delta\rho/\rho$ should be of order $10^{-5}$.
However, since inflatons of first-order inflation models, including ours,
are not slow-rolling fields,
the models need an additional scalar field
to generate the appropriate density perturbation.
Any scalar field with mass $m\ll H$ has
a quantum fluctuation of amplitude $H/2\pi$
at horizon crossing during inflation\cite{desitter}.
In the extended inflation model a Brans-Dicke field
takes a role of the extra scalar field, and
some of two fields inflation models\cite{twofields} have
a similar additional field.
So it seems not so unnatural to consider another scalar field to generate
the density fluctuation. 

Furthermore, it is also possible to produce scale invariant density
fluctuation even without rolling fields. 
The energy density of cosmic string \cite{string}
at time $\tau$ is $\rho_s\sim \mu \tau^{-2}\sim T^2_s
\tau^{-2}$, where $\mu$ is the energy per unit  length and
$T_s$ is the temperature of the string formation.
Since the radiation energy density $\rho$ evolves
as $1/G\tau^{2}$, any scale invariant perturbation should
be proportional to $\tau^{-2}$.
Hence, the cosmic string with $T_s\sim 10^{16}~GeV$
could explain the observed distortion of cosmic background 
radiation\cite{stringdensity}, as is well known.

In summary, invoking the renormalization 
group effects in the first-order inflation model
could resolve  the graceful-exit problem   in the general relativity context
with  quantum bubble nucleation. 
In this model a successful inflation requires the group theoretical restrictions
on the inflaton field. This inflation model may also constraint
the gauge group of GUTs, if the inflaton is a GUTs Higgs scalar.

\bigskip

 Authors  are thankful to J. C. Lee and J. S. Lee for helpful discussions. 
 This work was supported in part by KOSEF.

\end{document}